\documentclass[runningheads]{llncs}
\usepackage[T1]{fontenc}
\usepackage{graphicx}
\usepackage{hyperref}
\begin{document}
\title{From Education to Evidence: A Collaborative Practice Research Platform for AI-Integrated Agile Development}
\titlerunning{From Education to Evidence}
%
\author{ 
Tobias Geger\inst{1}\orcidID{0009-0004-4469-534X} \and
Andreas Rausch\inst{1}\orcidID{0000-0002-6850-6409} \and
Ina Schiering\inst{2}\orcidID{0000-0002-7864-5437} \and
Frauke Stenzel\inst{1}
\and
Stefan Wittek\inst{1}\orcidID{0009-0007-3877-625X}
} 
\authorrunning{T. Geger et al.}
%
\institute{Clausthal University of Technology, Clausthal-Zellerfeld, Germany
\email{\{thomas.tobias.marcello.geger, andreas.rausch, frauke.stenzel, stefan.wittek\}@tu-clausthal.de}\\
\and
Ostfalia University of Applied Sciences, Wolfenbüttel, Germany\\
\email{i.schiering@ostfalia.de}}
\maketitle              
\begin{abstract}
Agile software development evolves so rapidly that research struggles to remain timely and transferable—an issue heightened by the swift adoption of generative AI and agentic tools. Earlier discussions highlight theory and time gaps, leading to results that often lack clear reuse conditions or arrive too late for practical decisions. This paper introduces a project-based, AI-integrated agile education platform as a collaborative research environment, positioned between controlled studies and real-world industry. The platform enables rapid inquiry through sprint rhythms, quality gates, and genuine stakeholder involvement. We present a framework specifying, iteration structures, recurring events, and quality gates for AI-assisted engineering artifacts. Early results from several semesters—covering project pipeline, cohort growth, and stakeholder participation—show the platform’s potential to generate practice-relevant evidence efficiently and with reusable context. Finally, we outline future steps to enhance governance and evidence capture.

\end{abstract}
\section{Introduction}

Agile software development \cite{beck2001agile} evolves rapidly, driven by distributed work, new tool ecosystems, and the rise of AI-supported development. This often leaves research lagging behind, with practitioners viewing academic results as too slow or impractical. Practice research aims to close this gap, but major challenges remain, notably the theory gap—where results lack clear rationale and transferability—and the time gap—where industry changes faster than research cycles, making findings arrive too late. Both gaps hinder the transfer of relevant, reusable insights \cite{neumann2025closercollaborationpracticeresearch}.

The rapid adoption of generative AI and agentic tools makes these challenges even more urgent. AI now influences not only how teams develop and test software, but also shapes core agile practices. This raises questions around quality, responsibility, and trust, increasing the need for timely and transferable socio-technical guidance.

Recent discussions highlight two main research strategies: research-led (Mode-1) studies using controlled, theory-driven designs, and collaborative practice research platforms (Mode-2), where researchers and practitioners co-create and test interventions in real settings, sharing insights through rapid feedback loops \cite{neumann2025closercollaborationpracticeresearch}. The latter is especially suitable for the fast-moving context of AI-integrated agile development.

We propose to realize such a collaborative platform via project-based, AI-integrated agile education. Structured sprint cycles, explicit quality gates, and authentic stakeholder involvement position this approach between lab studies and industry practice, enabling rapid, contextualized research.

The study program becomes a long-term collaboration space where practitioners contribute real challenges and students test solutions in sprint-based teams. This supports fast iteration, systematic evidence capture, and the packaging of transferable guidance for practice.

Our investigation is structured around three questions: (1) How can project-based, AI-integrated agile education serve as an effective collaborative research platform given these challenges? (2) How can this be implemented through a practical framework with, sprints, and quality gates? (3) What early results and implications emerge for timely, transferable AI-integrated agile practice?

The paper is structured as follows: \autoref{Study Program} presents the conceptual framework and requirements; \autoref{Case Study} introduces the AI-integrated agile project framework; \autoref{Findings} summarizes early findings; \autoref{Conclusion} discusses limitations, scaling, and conclusions.

\section{Designing AI-Integrated Agile Education: AI Engineering Study Program} \label{Study Program}

Teaching agile effectively requires learning formats that go beyond lecture-centric instruction: students need repeated, authentic project settings in which teamwork, evolving requirements, and continuous feedback are experienced rather than explained. At TU Clausthal, we have already operationalized this principle in the project-based study program Digital Technologies, where agile methods (e.g., Scrum), project management, creativity techniques, and teamwork are taught and internalized through practice in real projects \cite{digitec}, \cite{StenzelKuepperRauschetal.2025}.

With generative AI, the need for such practice-oriented formats becomes even more pressing. LLMs and agentic AI increasingly support engineering activities, but they also introduce risks regarding quality, traceability, and responsible decision-making. Many curricula still implicitly prepare students for a workplace without AI, creating a qualification gap. We therefore follow the same core educational hypothesis that proved successful for agile education: project-based learning is a powerful vehicle to develop AI competence in context, because it forces students to apply AI tools under real constraints, to reflect on limitations and failure modes, and to take responsibility for outcomes through human oversight.

Building on these foundations, we developed the AI Engineering study program concept. Its key idea is to combine a shared core curriculum with discipline-specific specialization tracks. The core curriculum establishes a common baseline across all students---covering programming and software foundations, AI literacy and data-related skills, and agile development practices---so that students can collaborate in interdisciplinary settings. On top of this base, students choose an engineering focus (e.g., AI Software Engineering, AI Mechanical Engineering, AI Material Engineering), in which domain-specific competencies are developed and repeatedly connected to the use of AI as an engineering tool (``AI as a tool for engineering,'' not only ``AI as a product'').

\begin{figure}
    \centering
    \includegraphics[width=1\linewidth]{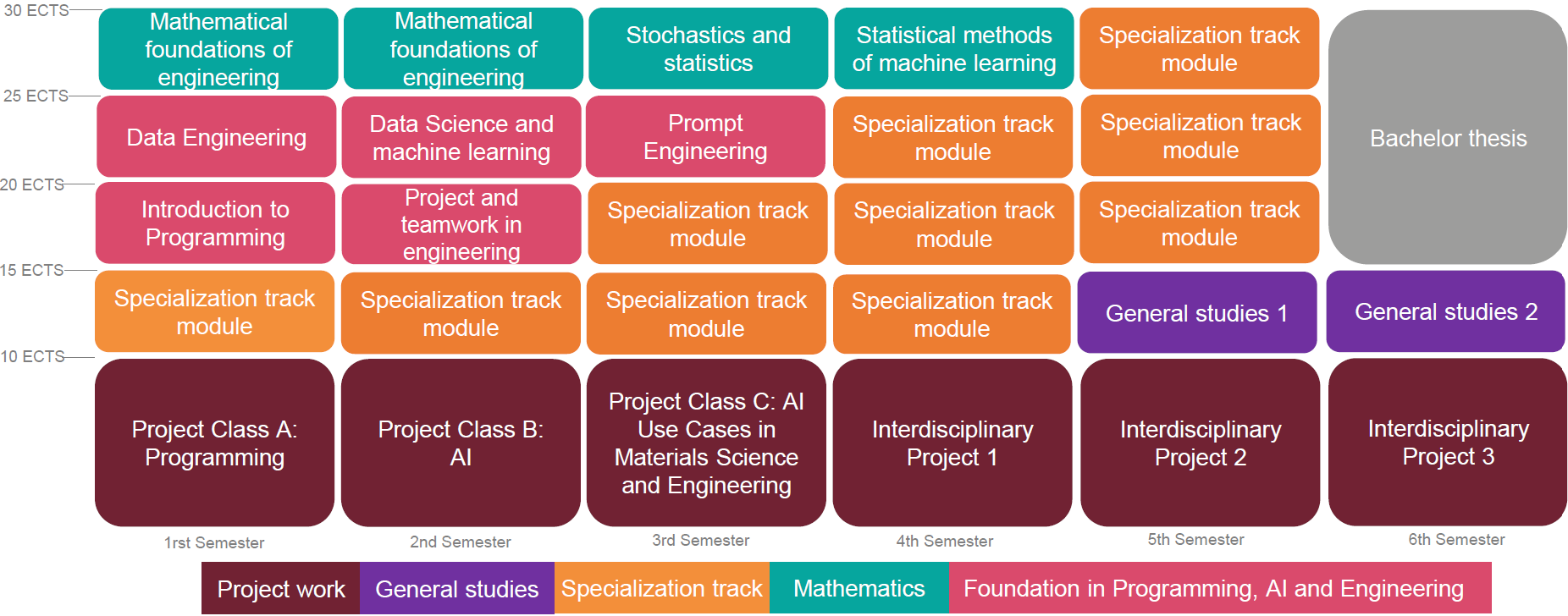}
    \caption{Model curricula of the bachelor program AI Engineering}
    \label{fig:curricula}
\end{figure}

A distinctive structural element of the program is that a substantial share of the workload is realized in project-oriented formats (approximately one third), where students apply agile methods and AI tools end-to-end on realistic problems: iteratively, in teams, and with continuous feedback. These projects are designed as authentic socio-technical learning settings in which students experiment with AI-supported workflows, evaluate results critically, and learn how to embed AI usage into engineering quality assurance and accountability. The curriculum itself is treated as a learning, agile system that is iteratively improved with input from students, educators, industry, and societal stakeholders, and designed for transfer (e.g., reusable materials and modular extensions).

The structure of the study program is summarized in the model curricula shown in Figure \ref{fig:curricula}: they make explicit (i) the shared core curriculum that establishes a common foundation in programming, mathematics, AI/data competencies, and agile ways of working, and (ii) the subsequent engineering specialization tracks that deepen domain-specific knowledge while maintaining a consistent AI-augmented engineering perspective. Importantly, the figures also highlight the program’s strong project-based backbone: project classes and recurring interdisciplinary projects are distributed across the semesters to ensure that students repeatedly apply agile practices and AI tools in authentic team settings---first to build fundamental skills, then to develop and evaluate discipline-specific AI use cases, and finally to integrate them in larger interdisciplinary projects and the bachelor thesis.


\paragraph{From study program to collaborative practice research platform.} Beyond education, the program functions as a collaborative practice research platform for AI‑integrated agile development, linking researchers (frame questions, synthesize insights), practitioners and external stakeholders (bring real problems, constraints, success criteria), and students (implement and evaluate sprint‑based interventions under realistic constraints). Evidence is captured via a recurring artifact backbone (backlogs/user stories, repositories, architecture/test artifacts, reviews, retrospectives) and explicit quality gates (review parties, repository‑based assessments, oral exams) to ensure accountability for AI‑assisted engineering. This setup enables timely, context‑rich, transferable evidence and underpins the early findings in \autoref{Findings}.

\section{AI-Integrated Agile Project Framework: Case Study} \label{Case Study}

This section reports a case study of our second-semester bachelor projects in which teams conduct \emph{AI-integrated} agile development, i.e., they deliberately use AI tools as part of their everyday engineering work while remaining accountable for outcomes. The following subsections operationalize our AI-Integrated Agile Project Framework through roles, iteration structures, learning mechanisms, and quality gates. Students enter with basic skills in Python, knowledge of object-oriented programming in Java, and familiarity with fundamental development tools such as Git. In addition, foundations in UML-based modeling, Scrum-based development, and project management have been established in prior courses. This project serves as a repesentative example of the projects conducted in later semesters, since all structural elementes, and even time schedules are consistent between all projects.

\subsection{Roles and responsibilities (including AI accountability)}
Teams typically consist of four to six developers, a Product Owner (teaching assistant), and a Supervisor (professor). While students initially act as developers, they gradually assume Product Owner responsibilities, for instance by writing user stories and refining acceptance criteria. The technical goal of the project is to develop a browser-based multiplayer game based on a given physical board game \cite{hof}. In AI-integrated work, a key expectation is accountability: teams must be able to explain and justify AI-supported decisions and artifacts (e.g., code, tests, and design choices), including limitations and risks.

\subsection{Iteration rhythm and events}
The project follows the Scrum framework over seven two-week sprints. Standard events include bi-weekly Dailies, mid-sprint ``InBetween'' presentations, and Sprint Reviews. Additionally, cross-team ``Review Parties'' are held after sprints 3, 5, and 7 to present results to a wider audience and to create comparable checkpoints across teams.

\begin{figure}
    \centering
    \includegraphics[width=1\linewidth]{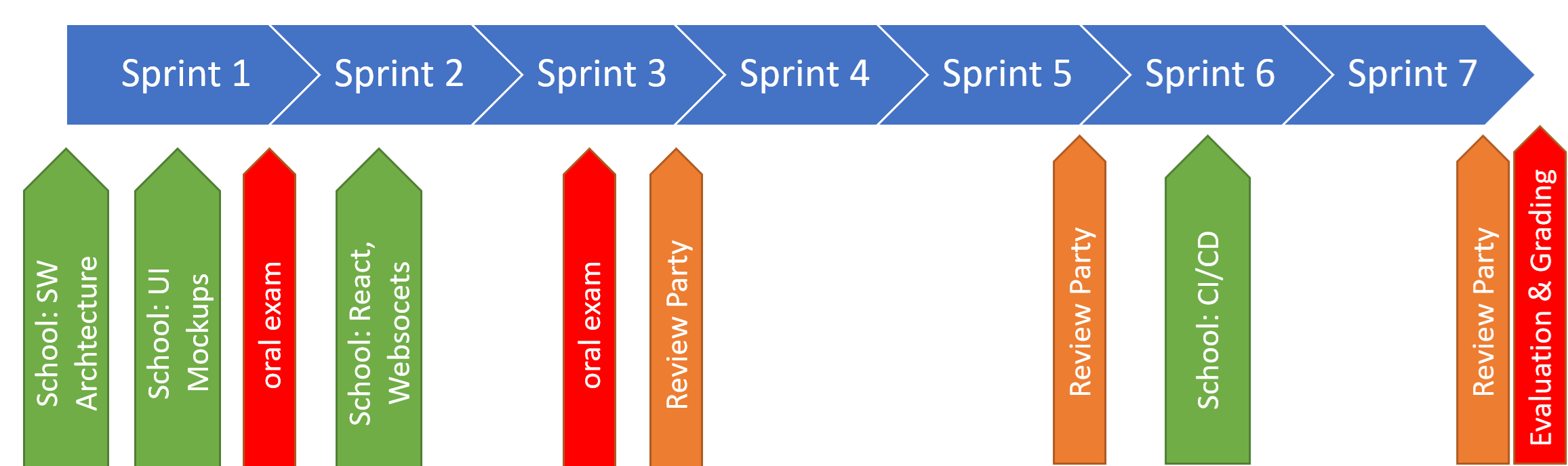}
    \caption{Time schedule of AI-integrated Agile Projects}
    \label{fig:schedule}
\end{figure}

Figure \ref{fig:schedule} summarizes the time schedule of a typical project, including the integration of recurring learning units (Schools) into the sprint cadence.

\subsection{Learning and transfer mechanisms}
To introduce new knowledge during the project, we employ ``Schools''---hybrid lecture-exercise units combined with homework and clustered into two units: technology and architecture. The technology unit covers topics such as React or TypeScript, while the architecture unit focuses on concepts such as client-server architecture and appropriate modeling techniques. AI assistance is not treated as a separate topic but embedded into each School: instructors demonstrate AI capabilities and limitations alongside the technical introduction (e.g., using AI for feedback, code explanation, or exploring implementation alternatives), and assignments encourage students to use AI tools critically (e.g., verification, refactoring, and documentation).

\subsection{Quality gates and evidence points}
While AI tools can be used throughout the project, we ensure that students maintain a deep understanding of their work through explicit quality gates. Students undergo individual oral exams (20 minutes) in which they explain their code and answer questions regarding implementation decisions and their use of AI. In addition, teams maintain a presentation that also serves as living documentation of requirements, architecture, process, and AI usage practices. The project grade is derived from the repository and this documentation, and individual grades are adjusted by peer ratings collected in a guided final session. These quality gates serve a dual purpose: assessment in education and recurring evidence points for practice research in AI-integrated agile development.

\section{Early Findings: Observations and Implications} \label{Findings} 

This section reports early evidence from applying the proposed collaborative practice research platform in the Digital Technologies/AI Engineering context. Our goal is not to claim generalizable effects, but to surface recurring observations that (i) indicate feasibility of the platform approach at scale, (ii) highlight where it helps to mitigate timeliness and transfer challenges in AI-integrated agile work, and (iii) form implications for future research and platform refinement.

\subsection{Evidence sources and scope}
\label{subsec:evidence-sources}
We triangulate early evidence across (a) project pipeline statistics collected per semester (e.g., number of pitched vs.\ conducted projects, number of participating students, industry vs.\ in-house projects, and typical team sizes), (b) cohort and intake development indicators, and (c) the recurring project rhythm and produced artifacts (e.g., review presentations, repositories, and additional project-specific documentation). In addition, we consider qualitative signals that become visible through repeated agile events and quality gates (e.g., review parties, retrospectives, and individual oral exams) as structured reflection points that continuously reveal recurring challenges and improvement opportunities.

\subsection{Quantitative snapshot: project pipeline, stakeholder involvement, and growth}
\label{subsec:quant-snapshot}
Across the observed semesters, the platform sustained a stable project throughput while cohort sizes increased. In the period from summer semester~2023 to winter semester~2025, the number of conducted projects grew from 7 to 18 per semester, with typical team sizes ranging from 4 to 6 students. Over the same period, the number of students participating in projects increased from 26 to 111. The project pipeline is intentionally over-provisioned: for example, in winter semester~2025, 20 projects were pitched and 18 were conducted, enabling a selection mechanism aligned with student interest while maintaining sufficient capacity.

A key indicator for transfer-oriented, AI-integrated agile work is authentic stakeholder involvement. Industry (real-stakeholder) projects were continuously offered each semester, typically in the range of 1-3 projects, and covered diverse partners and domains (e.g., mobility, smart city, and AI-enabled systems). This recurring stakeholder participation provides a stable channel for bringing in current constraints, expectations, and success criteria, while keeping project themes connected to fast-moving practice needs.

The platform also evolved under clear growth pressure. Overall program size increased substantially over time, especially on the master's level, resulting in larger and more diverse cohorts. Such growth is important for practice research: it increases the opportunity for repetition and cross-team triangulation, and it creates a broader empirical base for identifying recurring patterns rather than one-off anecdotes.

\subsection{Observations and implications}
\label{subsec:observations}

\textbf{O1: A sprint-based project platform can scale while preserving repeatability.}
Even with more students and projects, the platform maintained a regular process, enabling rapid, comparable research cycles. 
\emph{Implication:} Favor regular sprint rhythms and checkpoints over ad-hoc structures for faster feedback and comparability.

\textbf{O2: Over-provisioning (pitched vs.\ conducted projects) acts as a built-in selection and relevance mechanism.}
More projects are pitched than conducted; only popular or relevant ones proceed, supporting timely validation of emerging topics. 
\emph{Implication:} Use project selection as a relevance signal to keep research current.

\textbf{O3: Continuous real-stakeholder participation strengthens transfer by making context explicit from the outset.}
Industry involvement gives early clarity on constraints and criteria, which helps make findings reusable.
\emph{Implication:} Institutionalize stakeholder involvement as an essential part of the platform.

\textbf{O4: Quality gates create structured evidence points for AI-integrated work (beyond tool use).}
Quality gates require teams to clearly justify AI-driven decisions, focusing on accountability over mere tool use. 
\emph{Implication:} Design quality gates for traceability, explainability, and responsibility.

\textbf{O5: The recurring artifact set enables packaging for reuse and cumulative learning.}
Consistent artifacts each semester make capturing and transferring lessons learned easier. 
\emph{Implication:} Standardize a minimal artifact structure and add AI-specific documentation as needed.

\textbf{O6: Growth increases triangulation potential, but also demands stronger platform governance.}
Larger cohorts enable more cross-team insights, but also require better coordination and standards. 
\emph{Implication:} Apply explicit governance to maintain quality and comparability when scaling.

\subsection{Summary}
Overall, the early evidence suggests that project-based AI-integrated agile education can function as a collaborative practice research platform that supports rapid cycles of inquiry (timeliness) while improving transfer through stakeholder anchoring and systematic artifact packaging. At the same time, the observations highlight that scale and transfer do not happen automatically: they depend on explicit platform governance, consistent quality gates, and deliberate documentation practices tailored to AI-integrated engineering work.

\section{Discussion and Conclusion} \label{Conclusion}
Limitations of this study are that mainly qualitative measurements as observations in the context of review parties, retrospectives, and individual oral exams are employed. It would be interesting to investigate in addition quantitative measurements \cite{choras2020softwaremetrics} in the context of the practice research platform. Also it would be interesting to investigate the potential to built up an agile mindset and agile culture \cite{neumann2024agilecultureclashunveiling} incorporating respect, trust, learning culture and feedback culture.

With this case study including observations and implications an approach for a collaborative practice research platforms (Mode-2), for co-creating and testing interventions in real settings and rapid feedback loops  \cite{neumann2025closercollaborationpracticeresearch} is presented. Key challenges for scaling the platform are to preserve relevance of projects and authentic stakeholder integration while a strong platform governance to maintain quality and comparability is established. Based on our experiences we are interested in fostering the participation of stakeholders from research and practice in the development and evaluation of this collaborative practice research platforms.

\begin{credits}
\subsubsection{\ackname} 
This study was funded by the Lower Saxony Ministry for Science and Culture (grant number 11 – 76251-1900/2021 and 11 – 76251-1899/2021).

\end{credits}

%
%
\bibliographystyle{splncs04}
\bibliography{mybibliography}

@inproceedings{StenzelKuepperRauschetal.2025,
  author    = {Stenzel, Frauke and K{\"u}pper, Steffen and Rausch, Andreas and Schiering, Ina and Bikker, Gert},
  title     = {Projektbasierte {L}ehre in der {I}nformatik zur {E}ntwicklung von {Z}ukunftskompetenzen},
  booktitle   = {Tagungsband zum 6. Mint Symposium: Zukunft MINT Lehre: Was bleibt? Was kommt? Was wirkt?},
  editor    = {D{\"o}lling, Hanna and Sch{\"a}fle, Claudia},
  publisher = {BayZiel},
  address   = {M{\"u}nchen},
  doi       = {10.57825/repo_in-6376},
  pages     = {170 -- 177},
  year      = {2025},
  language  = {de}
}

@online{digitec,
    author = {{Center for Digital Technologies}},
    title = {{D}igital {T}echnologies},
    year = {2026},
    url = {https://digitecstudieren.de},
    note = {accessed: 2026-01-20}
}

@online{hof,
    author = {{Center for Digital Technologies}},
    title = {{Zweitsemesterprojekte}},
    year = {2026},
    url = {https://projects.digitecstudieren.de},
    note = {accessed: 2026-01-20}
}

@misc{neumann2025closercollaborationpracticeresearch,
      title={Towards a Closer Collaboration Between Practice and Research in Agile Software Development Workshop: A Summary and Research Agenda}, 
      author={Michael Neumann and Eva-Maria Schön and Mali Senapathi and Maria Rauschenberger and Tiago Silva da Silva},
      year={2025},
      eprint={2507.10785},
      archivePrefix={arXiv},
      primaryClass={cs.SE},
      url={https://arxiv.org/abs/2507.10785}, 
}

@ARTICLE{choras2020softwaremetrics,
  author={Choraś, Michał and Springer, Tomasz and Kozik, Rafał and López, Lidia and Martínez-Fernández, Silverio and Ram, Prabhat and Rodriguez, Pilar and Franch, Xavier},
  journal={IEEE Access}, 
  title={Measuring and Improving Agile Processes in a Small-Size Software Development Company}, 
  year={2020},
  volume={8},
  number={},
  pages={78452-78466},
  doi={10.1109/ACCESS.2020.2990117}}

@misc{beck2001agile,
  title={Agile manifesto. Softw Dev},
  author={Beck, K and Beedle, M and Van Bennekum, A and Cockburn, A and Cunningham, W and Fowler, M and Thomas, D},
  doi={doi. org/10.1177},
  volume={405},
  pages={411},
  year={2001}
}

@misc{neumann2024agilecultureclashunveiling,
      title={Agile Culture Clash: Unveiling Challenges in Cultivating an Agile Mindset in Organizations}, 
      author={Michael Neumann and Thorben Kuchel and Philipp Diebold and Eva-Maria Schön},
      year={2024},
      eprint={2405.15066},
      archivePrefix={arXiv},
      primaryClass={cs.SE},
      url={https://arxiv.org/abs/2405.15066}, 
}
%




\end{document}